\documentclass[11pt]{article}
\title{On Resolvent Identities in Gaussian Ensembles at the Edge of the Spectrum}
\author{Alexander Soshnikov \thanks{
Department of Mathematics,
University of California at Davis, 
One Shields Ave., Davis, CA 95616, USA.
E-mail address: soshniko@math.ucdavis.edu.
Research was supported in part by the 
NSF grant DMS-0707145. 
}  }
\usepackage{epsfig}
\usepackage{amsfonts}
\usepackage{amssymb}
\usepackage{amsthm}
\usepackage{amstext}
\usepackage{amscd}
\usepackage{amsmath}
\usepackage{delarray}
\begin{document}
\maketitle
\newtheorem{theo}{Theorem}[section]
\newtheorem{prop}{Proposition}[section]
\newtheorem{lemme}{Lemma}[section]
\newtheorem{conjecture}{Conjecture}[section]
\newtheorem{definition}{Definition}[section]
\newtheorem{fact}{Fact}[section]
\newtheorem{hyp}{Assumption}[section]
\theoremstyle{remark}
\newtheorem{rem}{Remark}
\newtheorem{remark}{Remark}[section]
\newtheorem{Remark}{Remark}[section]
\newtheorem{Notationnal remark}{Remark}[section]
\newcommand{\bremnot}{\begin{Notationnal remark}}
\newcommand{\eremnot}{\end{Notationnal remark}}
\newcommand{\brem}{\begin{remark}}
\newcommand{\erem}{\end{remark}}
\newcommand{\bconj}{\begin{conjecture}}
\newcommand{\econj}{\end{conjecture}}
\newcommand{\bdefi}{\begin{definition}}
\newcommand{\edefi}{\end{definition}}
\newcommand{\bt}{\begin{theo}}
\newcommand{\bfa}{\begin{fact}}
\newcommand{\efa}{\end{fact}}
\newcommand{\Si}{\Sigma}
\newcommand{\mbE}{\mathbb{E}}
\newcommand{\mL}{\mathcal{L}}
\newcommand{\mP}{\mathcal{P}}
\newcommand{\mQ}{\mathcal{Q}}
\newcommand{\mR}{\mathcal{R}}
\newcommand{\et}{\end{theo}}
\newcommand{\bp}{\begin{prop}}
\newcommand{\ep}{\end{prop}}
\newcommand{\bl}{\begin{lemme}}
\newcommand{\el}{\end{lemme}}
\newcommand{\be}{\begin{equation}}
\newcommand{\ee}{\end{equation}}
\newcommand{\bmp}{\begin{pmatrix}}
\newcommand{\emp}{\end{pmatrix}}
\newcolumntype{L}{>{$}l<{$}}
\newenvironment{Cases}{\begin{array}\{{lL.}}{\end{array}}
\begin{abstract}
We obtain the recursive identities for the joint moments of the traces of the powers of the resolvent
for Gaussian ensembles of random matrices at the soft and hard edges of the spectrum.  We also discuss the possible ways to extend these 
results to the non-Gaussian case.
\end{abstract}

\section{Introduction}
Consider the Gaussian Orthogonal Ensemble (GOE) of real symmetric $ n\times n $ random matrices
\begin{equation}
A_n=\frac{1}{\sqrt n}\left ( a_{ij}\right)_{i,j=1}^n,
\end{equation}
where $ \{a_{ij}=a_{ji} \}_{i\leq j} $ are independent $N (0, 1+\delta_{ij}) $ random variables.
GOE is the archetypal example of a Wigner real symmetric random matrix where the matrix entries 
$ \{a_{ij}=a_{ji} \}_{i\leq j} $ are assumed to be independent up from the diagonal, centralized, and to have the same 
variance (except, possibly, on the diagonal).  It follows from the classical Wigner semi-circle law 
((\cite{Wig1}, \cite{Wig2}, \cite{Arn}) that the empirical distribution 
function of the eigenvalues of $A_n$ converges as $ n \to \infty $ to the limiting distribution 
with the probability density $ \frac{1}{2\*\pi}\*\sqrt{4-x^2} $  supported on the interval $[-2, 2]. $ Celebrated work by Tracy and Widom 
(see \cite{TW1} for the GOE case) proved that the largest eigenvalues of $A_n$ deviate from the right edge of the spectrum
on the order of $n^{-2/3}. $  In particular, Tracy and Widom calculated the limiting distribution of the largest eigenvalue:
\begin{equation}
\label{tw1}
\lim_{n\to \infty} \mathbb{P} \left( \lambda_{max} \leq 2 +  x \*n^{-2/3} \right)= F_1(x)=
\exp\left(-1/2\*\int_x^{\infty} q(t)+(t-x)\*q^2(t)\*dt \right), 
\end{equation} 
where $q(x) $ is the solution of the Painl\'{e}ve II differential equation $ \ q''(x)=x\*q(x) +2\*q^3(x) \ $ with the 
asymptotics  at infinity
$ q(x) \sim Ai(x) $ as $ x \to +\infty. \ $   Here $Ai(x)$ denotes the Airy function.

To consider the joint distribution of the largest eigenvalues at the edge of the spectrum,
we rescale the eigenvalues as
\begin{equation}
\label{rescaling}
\lambda^{(n)}_j= 2 + \xi^{(n)}_j \* n^{-2/3}, \ \ j=1,2, \ldots,n.
\end{equation}
where $ \lambda^{(n)}_1 \geq  \lambda^{(n)}_2 \ldots \geq \lambda_n^{(n)} $ are the ordered eigenvalues of $A_n.$
It then follows from the results of \cite{So2}, \cite{TW3}  that the random point configuration 
$ \{\xi^{(n)}_j, \ \ 1\leq j\leq n \} $ converges in distribution on the cylinder sets to the random point process on the real line
with the $k$-point correlation functions given by
\begin{equation}
\label{pfaffcor}
\rho_k(x_1, \ldots, x_k)=\left(\det \left(K(x_i, x_j) \right)_{1\leq i,j \leq k} \right)^{1/2},
\end{equation}
where $K(x,y)$ is a $2 \times 2$ matrix-valued kernel with the entries
\begin{eqnarray}
\label{matrixkernel}
& & K_{11}(x,y)= K_{Airy}(x,y)+ \frac{1}{2} Ai(x) \* \left(1-\int_y^{+\infty} Ai(z) \* dz \right), \\
& & K_{12}(x,y)= - \partial_y K_{Airy}(x,y) - \frac{1}{2}\*Ai(x)\* Ai(y), \\
& & K_{21}(x,y)=-\int_x^{+\infty} K_{Airy}(z,y)\*dz + \frac{1}{2} \left(\int_y^x Ai(z)\* dz + \int_x^{+\infty} Ai(z) \* dz
\* \int_y^{+\infty} Ai(z) \* dz \right), \\
& & K_{22}(x,y)=K_{11}(y,x),
\end{eqnarray}
and the Airy kernel $K_{Airy}(x,y)$ is defined as
\begin{equation}
\label{airykernel}
K_{Airy}(x,y)= \int_0^{+\infty} Ai(x+z)\* Ai(y+z) \* dz = \frac{Ai(x)\* Ai'(y) - Ai'(x)\* Ai(y)}{x-y}.
\end{equation}
Therefore, the $k$-point correlation function of the limiting random point process is given by the square root of the 
determinant of the $2k \times 2k $ matrix defined in (\ref{pfaffcor}) and (\ref{matrixkernel}).  One can also rewrite the 
$k$-point correlation function in the pfaffian form (see e.g. \cite{Mehta}) which shows that the limiting random point 
process belongs to the family of the pfaffian random point processes (see e.g. \cite{So3}).
In particular, the right-most particle of this pfaffian random point process is given by the Tracy-Widom distribution
(\ref{tw1}).  Moreover, it was shown in \cite{So1} that the asymptotic  behavior of the largest eigenvalues
is universal for Wigner real symmetric matrices with sub-Gaussian and symmetrically distributed entries.  

Define 
\begin{equation}
\label{resolvent}
G_n(z)=(A_n-2-z\*n^{-2/3})^{-1} 
\end{equation}
for complex $z$ with non-zero imaginary part $ \Im z \not=0.$  Here and throughout the paper, we will use $ (A_n-z)^{-1} $ as the shorthand notation
for the resolvent matrix $ (A_n-z\*Id)^{-1}.$ 

Let
\begin{equation}
\label{stepenres}
g_{n,k}(z)= n^{-2\*k/3} \* {\text  Tr} G_n^k(z) =n^{-2\*k/3}{\text  Tr} (A_n-2-z\*n^{-2/3})^{-k} = \sum_1^n 
(\xi_j^{(n)} -z)^{-k}
\end{equation}
for positive integers $k=1,2,\ldots.$
It can be shown that for $k \geq 2,  \ g_{n,k}(z) $ is a ``local'' statistic of the largest eigenvalues in the GOE.  Indeed, only eigenvalues
from the $O(n^{-2/3})$-neighborhood of the right edge of the spectrum give non-vanishing contribution to
$ g_{n,k}(z)$ in the limit $ n \to \infty. $  For example, the joint contribution of the eigenvalues from 
$ (-\infty, 2-\delta] $ can be trivially bounded in absolute value by $n^{1-2\*k/3} \* |\delta + z\*n^{-2/3}|^k =o(1) $
for large $n$ uniformly in $z$ with $\Re z $ bounded from below.   More delicate estimates involving the asymptotics of the one-point correlation 
function, imply that the joint contribution of the eigenvalues from $ (-\infty, 2-n^{-2/3+\varepsilon}) $ to $g_{n,k}(z)$ is still negligible
for all $\varepsilon >0 $ and $k >1.$  Moreover, the one-point correlation function $\rho_1(x) $
of the limiting pfaffian random point process defined in (\ref{rescaling})-(\ref{airykernel}) decays super-exponentially
at $+\infty $ and grows proportionally to $|x|^{1/2} $ at $-\infty. $   Consequently,  if $\xi= \{\xi_j, 
\ j \in \mathbb{Z} \} $ is a random point configuration of the limiting pfaffian random process then
\begin{equation}
\label{integral}
\mbE  \sum_j  |\xi_j -z|^{-k} = \int_{-\infty}^{+\infty} |x-z|^{-k} \* \rho_1(x) \* dx <\infty 
\end{equation}
for any integer $k \geq 2. $
The integral at the r.h.s. of (\ref{integral}) diverges for $k=1$ which emphasizes the fact that $g_{n,1}(z) $ is not a 
``local statistic'' as the main contribution to $g_{n,1}(z)=n^{-2/3} \*\times $ \\
${\text  Tr} (A_n-2-z\*n^{-2/3})^{-1} $ comes from the 
eigenvalues in the bulk.  Moreover, it could be shown from the asymptotics of the GOE one-point correlation function that
\begin{equation}
\label{matozh}
\mbE \left({\text  Tr} (A_n-2-z\*n^{-2/3})^{-1} \right)= -n +O(n^{2/3}).
\end{equation}
Eventhough   the eigenvalues from the bulk of the spectrum give the main contribution to the mathematical expectation
of ${\text  Tr} (A_n-2-z\*n^{-2/3})^{-1}, $ their joint 
contribution to the fluctuations of ${\text  Tr} (A_n-2-z\*n^{-2/3})^{-1} $
around its mean is much smaller (namely, it can be shown to be of order of constant if one smoothes their contribution by a test function with the 
support inside $[-2+\delta, 2\\delta]. $)   On the other hand, the largest eigenvalues of $A_n$ give smaller 
(namely, of the order of $O(n^{2/3})$) contribution to the mean of ${\text  Tr} (A_n-2-z\*n^{-2/3})^{-1}, $ but they
give the main contribution to the fluctuations of ${\text  Tr} (A_n-2-z\*n^{-2/3})^{-1} $ around its mean.
This suggests to consider 
\begin{equation}
\label{centralized}
g_{n,1}^c(z)= n^{-2/3} \* \left( n + {\text  Tr} G_n(z)\right) =n^{-2/3}\left(n+{\text  Tr} (A_n-2-z\*n^{-2/3})^{-1} \right)
\end{equation}
which is a ``local'' statistic in a sense that the main contribution to $g_{n,1}^c(z)$ comes from the largest eigenvalues
(i.e. the eigenvalues that deviate from the right edge of the spectrum on the order of $O(n^{-2/3}).$)

In Theorem 1.1, we obtain the recursive relations on the joint moments of the local linear statistics $g_{n,1}^c(z) $
and $ g_{n,k}(z), \ \ k\geq 2. \ $   Let $K$ be a multi-index, $K=(k_1, \ldots, k_j), \ \ j \geq 1, $ with the components
$k_l, \ 1\leq l \leq j, $ nonnegative integers.  The number of components $j$ is not fixed.  
We will denote by $m_K $ the corresponding joint moment of  $g_{n,1}^c(z) $
and $ g_{n,k}(z), \ \ k\geq 2, $  namely:
\begin{equation}
\label{jointmoment}
m_K= \mbE \left( (g_{n,1}^c(z))^{k_1} \* \prod_{l=2}^j  (g_{n,l}(z))^{k_l} \right).
\end{equation}
Let $e_l$ denote the multi-index with the $l$-th component equal to $1$ and the other components equal to zero.

\bt \label{mt}
Let $K$ be a non-zero multi-index, then the following equation holds:
\begin{eqnarray}
\label{recursiveGOE}
& & m_K \* (z +O(n^{-2/3})) - m_{K+2\*e_1} \* (1 + O(n^{-2/3})) - m_{K+e_2}\* (1 + O(n^{-2/3}))   \nonumber \\
& -& 2 \* \sum_{l\geq 1} l\*k_l \* m_{K-e_l+e_{l+2}} \*
(1+O(n^{-2/3})) = O(n^{-1/3}) \* m_{K+e_1}.
\end{eqnarray}
Also, the following ``boundary'' condition holds:
\begin{equation}
\label{recnule}
z +O(n^{-2/3}) - m_{2\*e_1}\* (1 + O(n^{-2/3})) - m_{e_2}\* (1 + O(n^{-2/3}))  =O(n^{-1/3}) \*m_{e_1}.
\end{equation}
\et

{\bf Remark.}  We will always assume in (\ref{recursiveGOE}) that
$ k_l \* m_{K-e_l+e_{l+2}} =0 $ if $k_l=0.$

Theorem 1.1 will be proved in the next section. Let us now consider the Gaussian Unitary Ensemble (GUE) of Hermitian $ n\times n $ random matrices.
\begin{equation}
A_n=\frac{1}{\sqrt n}\left ( a_{jk}\right)_{j,k=1}^n,
\end{equation}
where $ \{ \Re a_{jk}= \Re a_{kj} \}_{j< k} $ and $ \{ \Im a_{jk}= -\Im a_{kj} \}_{j< k} $
are i.i.d. $N (0, 1/2) $ random variables, and $ \{a_{ii} \}_{1\leq i \leq n} $ are i.i.d. $N(0,1)$ random variables.

The global distribution of the eigenvalues of $A_n$ still satisfies the Wigner semi-circle law in the limit $n \to \infty. $
The limiting local distribution of the largest eigenvalues of $A_n$ was calculated by Tracy and Widom in \cite{TW2}.  In 
particular,

\begin{equation}
\label{tw2}
\lim_{n\to \infty} \mathbb{P} \left( \lambda_{max} \leq 2 +  x \*n^{-2/3} \right)= F_2(x)=
\exp\left(-\int_x^{\infty}(t-x)\*q^2(t)\*dt \right), 
\end{equation} 
where, as before, $q(x) $ is the solution of the Painl\'{e}ve II differential equation 
with the same asymptotics  at infinity.

Consider the same rescaling at the right edge of the spectrum as in the GOE case, namely
\begin{equation}
\label{rescaling2}
\lambda^{(n)}_j= 2 + \xi^{(n)}_j \* n^{-2/3}, \ \ j=1,2, \ldots,
\end{equation}
where $ \lambda^{(n)}_1 \geq  \lambda^{(n)}_2 \ldots \geq \lambda_n^{(n)} $ are the ordered eigenvalues of $A_n.$
It then follows from the results of \cite{TW2} that the random point configuration 
$ \{\xi^{(n)}_j, \ \ 1\leq j\leq n \} $ converges in distribution on the cylinder sets to the random point process on the real line
with the $k-$point correlation functions given by
\begin{equation}
\label{detcor}
\rho_k(x_1, \ldots, x_k)=\det \left(K(x_i, x_j) \right)_{1\leq i,j \leq k} 
\end{equation}
where $K(x,y) = K_{Airy}(x,y)$ is the Airy kernel defined in (\ref{airykernel}).
The limiting random point process belongs to the class of determinantal random point processes
(see \cite{So4}, \cite{YPBV}).  

Let us use the same notations $G_n(z), \ g_{n,k}(z), \ g_{n,1}^c(z), $ and $ M_K $ in the GUE case as they were defined  in
(\ref{resolvent}), (\ref{stepenres}), (\ref{centralized}), and (\ref{jointmoment}) in the GOE case above.
The following analogue of the Theorem 1.1 holds:

\bt \label{MT}
Let $K$ be a non-zero multi-index, then the following equation holds:
\begin{eqnarray}
\label{recursiveGUE}
& & m_K \* (z +O(n^{-2/3})) - m_{K+2\*e_1} \* (1 + O(n^{-2/3}))  \nonumber \\
&-& \sum_{l\geq 1} \*l\*k_l \* m_{K-e_l+e_{l+2}} \*
(1+O(n^{-2/3})) = 
O(n^{-1/3}) \* m_{K+e_1}.
\end{eqnarray}
Also, the following ``boundary'' condition holds:
\begin{equation}
\label{recnule2}
z +O(n^{-2/3}) - m_{2\*e_1} \* (1 + O(n^{-2/3}))= O(n^{-1/3}) \* m_{e_1}.
\end{equation}
\et

We now turn our attention to Wishart (a.k.a Laguerre) ensembles of random matrices.  Again, we start with the real case.
Let $A=A_{n,N}= \frac{1}{\sqrt{n}} \* (a_{ij}) $ be a rectangular $n\times N$ matrix 
with $\{a_{ij}, \ 1\leq i \leq n, \ 1\leq j \leq N, \} $ real i.i.d. $N(0,1) $ random variables.
Let us assume that $N \geq n, \ $ and $ N-n=\nu $ is fixed.  Consider a nonnegative-definite random  
matrix 
\begin{equation}
\label{realWishart}
M_{n,N}= A\*A^t. 
\end{equation}
The ensemble of random matrices $M_{n,N}$ is known as the real Wishart distribution in the statistical literature or 
the Laguerre ensemble in the mathematical physics.    The empirical distribution function of the eigenvalues of
$M_{n,N} $ converges to the Marchenko-Pastur law as $n \to \infty $ (\cite{MP}, \cite{SB}).   
The density of the Marchenko-Pastur law is given by
\begin{equation}
\label{MP}
\rho_{MP}(x)=\left \{\begin{array} {r@{\quad:\quad}l} 
\frac{1}{2\*\pi\*\sqrt{x}} \*\sqrt{4-x}& \ \text{if} \ \ 0\leq x \leq 4, \\
0 & \text{otherwise}. \end{array} \right.
\end{equation}

Our goal is to study the distribution of the smallest  eigenvalues of $M_{n,N}$ in the limit \\ 
$n \to \infty, \  N-n=\nu. $
It can be shown (see e.g. \cite{F1}, \cite{F2}, \cite{E}) that the smallest eigenvalue of $M_{n,N}$ are of the order of
$n^{-2}. $  Moreover, if we consider the rescaling at the hard edge of the spectrum
\begin{equation}
\label{rescalinghard}
\lambda_i^{(n,N)}=\frac{\xi_i^{(n,N)}}{4\*n^2}, \ \ 1\leq i \leq n, 
\end{equation}
one can show that the random point configuration $ \{\xi_i^{(n)}, \ 1\leq i\leq n \} \ $ converges in distribution on the cylinder
sets to the pfaffian random point process on $(0, +\infty). \ $ The $k$-point correlation functions
of the limiting process are of the same form as in (\ref{pfaffcor}), where $K(x,y) $ is again a $2\times 2$ matrix-valued kernel.
The formulas for the entries of $K(x,y)$ are similar to (\ref{matrixkernel}) with the important difference being that the Airy
kernel $K_{Airy}(x,y)$ is replaced by the Bessel kernel
\begin{equation}
\label{Besselkernel}
K_{Bessel}(x,y)=\frac{ J_{\nu}(\sqrt{x})\*\sqrt{y}\*J_{\nu +1}(\sqrt{y}) - J_{\nu}(\sqrt{y})\*\sqrt{x}\*J_{\nu +1}(\sqrt{x})
}{2(x-y)}, 
\end{equation}
where $J_{\nu}(x) $ is the usual Bessel function of index $\nu. $

Define 
\begin{equation}
\label{lgov}
G_{n,N}(t)= \left(M_{n,N} +\frac{t^2}{n^2}\right)^{-1},
\end{equation}
where $t $ is a real number.  Then 
\begin{equation}
\label{urara}
g_k(t)=g_{n,N,k}(t)= n^{-2\*k} \* Tr G^k_{n,N}(t) =  
n^{-2\*k} \* Tr\left(M_{n,N} +\frac{t^2}{n^2}\right)^{-k}=\sum_1^n (\xi_i^{(n,N)}+t^2)^{-k} 
\end{equation}
is a ``local'' statistics for any positive integer $k=1,2,3, \ldots. $  Indeed, one can show that
the eigenvalues from the bulk of the spectrum give vanishing contribution to $g_k(t). $ In particular,
if $\xi=\{\xi_i, \ \ i\in \mathbb{N}  \} $ is a random point configuration of the limiting pfaffian process, then
\begin{equation}
\label{integral100}
\mbE  \sum_j  (\xi_j +t^2)^{-k} = \int_0^{+\infty} \frac{1}{(x+t^2)^k} \* \rho_1(x) \* dx <\infty.
\end{equation}
We are interested to study the joint moments of the linear statistics $ g_k(t), \ k\geq 1. \ $ Let, as before, 
$K=(k_1, \ldots, k_j), \ j\geq 1, $ denote a multi-index, and $m_K$ stand for the corresponding joint moment
\begin{equation}
\label{soccer}
m_K=  \mbE \prod_{l=1}^j  (g_{l}(t))^{k_l}.
\end{equation}
The following theorem holds.

\bt \label{MTeo}
Let $K$ be a non-zero multi-index, then the following equation holds:
\begin{eqnarray}
\label{recursiveWishartreal}
& & (\nu - t^{-2} + \frac{1}{n})\* m_{K+e_1} +m_{K+e_2} + m_{K+2\*e_1} 
+ 2 \* \sum_{l=1}^j l\*k_l \* m_{K-e_l+e_{l+2}} - \frac{2}{t^2} \* \sum_{l=1}^j l\*k_l \* m_{K-e_l+e_{l+1}} \cr
& & = \frac{1}{n}\*\frac{1}{t^2} \* m_K.
\end{eqnarray}
Also, the following ``boundary'' condition holds:
\begin{equation}
\label{recnule3}
(\nu - t^{-2} + \frac{1}{n})\* m_{e_1} +m_{e_2} + m_{2\*e_1}= t^{-2}.
\end{equation}
\et
We recall that $\nu=N-n \geq 0 $ is the difference between the dimensions of the rectangular matrix $A_{n,N}. $

We finish the Introduction by the discussion of the complex Wishart ensemble.
Let $A=A_{n,N}= \frac{1}{\sqrt{n}} \* (a_{ij}) $ be a rectangular $n\times N$ matrix with  $\{\Re a_{ij}, \Im a_{ij}, \ 1\leq i \leq n, \ 
1\leq j \leq N, \}$
i.i.d. $N(0,1/2) $ random variables.  As before, we us assume that $N \geq n, \ $ and $ N-n=\nu $ is fixed.  
Consider now a nonnegative-definite random matrix 
\begin{equation}
\label{complexWishart}
M_{n,N}= A\*A^*. 
\end{equation}
The ensemble of random matrices $M_{n,N}$ is known as the complex Wishart/Laguerre ensemble of random matrices.    
Consider the rescaling of the eigenvalues at the hard edge of the spectrum
\begin{equation}
\label{rescalinghard2}
\lambda_i^{(n,N)}=\frac{\xi_i^{(n,N)}}{4\*n^2}, \ \ 1\leq i \leq n, 
\end{equation}
It follows from the results of \cite{F1} that the random point configuration $ \{\xi_i^{(n)}, \ i\geq 1 \} \ $ converges in 
distribution on the cylinder sets to the determinantal random point process on $(0, +\infty) $ with the correlation 
kernel given by the  Bessel kernel (\ref{Besselkernel}) in the limit $ n \to \infty, \ N=n +\nu. $
Define $G_{n,N}, \ g_k(t), \ k\geq 1, $ and $m_K $ in the same way as in (\ref{lgov}), (\ref{urara}), and (\ref{soccer}).
The following theorem holds.

\bt \label{MTeor}
Let $K$ be a non-zero multi-index, then the following equation holds:
\begin{eqnarray}
\label{recursiveWishartcomplex}
& & (\nu + \frac{1}{n})\* m_{K+e_1} + m_{K+2\*e_1} 
+ \sum_{l=1}^j l\*k_l \* m_{K-e_l+e_{l+2}} - \frac{1}{t^2} \* \sum_{l=1}^j l\*k_l \* m_{K-e_l+e_{l+1}} \cr
& & = \frac{1}{n}\*\frac{1}{t^2} \* m_K.
\end{eqnarray}
Also, the following ``boundary'' condition holds:
\begin{equation}
\label{recnule4}
(\nu + \frac{1}{n})\* m_{e_1} +m_{2\*e_1}= t^{-2}.
\end{equation}
\et
We recall that $\nu=N-n \geq 0 $ is the difference between the dimensions of the rectangular matrix $A_{n,N}. $  Theorems 1.3 and 1.4 will be 
proved in Section 3.

\section{Proof of Theorems 1.1 and 1.2}

Let us start with the proof of Theorem 1.1.  Our first goal is to establish (\ref{recnule}).
To this end, we consider $n^{1/3}\* m_{e_1}= \mbE \left( n^{-1/3}\* \left(n + Tr G_n(z) \right) \right), $ where, as before,
$G_n(z)= (A_n-2-z\*n^{-2/3})^{-1}. \ $  By using the resolvent identity
\begin{eqnarray}
\label{rezolventy}
& & G_n(z)=(A_n-2-z\*n^{-2/3})^{-1}= -(2+z\*n^{-2/3})^{-1} \*Id + (2+z\*n^{-2/3})^{-1} \* A_n\* (A_n-2-z\*n^{-2/3})^{-1}= 
\nonumber \\
& & -(2+z\*n^{-2/3})^{-1} \*Id + (2+z\*n^{-2/3})^{-1} \* A_n\* G_n,
\end{eqnarray}
we arrive at
\begin{equation}
\label{moskva}
n^{1/3}\* m_{e_1}= n^{2/3} -(2+z\*n^{-2/3})^{-1} \* n^{2/3} +(2+z\*n^{-2/3})^{-1}\*n^{-1/3} \* \mbE \sum_{ij} A_{ij}\*G_{ji}.
\end{equation}
Here $A_{ij}=\frac{a_{ij}}{\sqrt{n}}$ denote the matrix entries of $A_n$, and $G_{ij}$ denote the matrix entries of $G_n(z).$
To calculate  $\mbE  A_{ij}\*G_{ji}, $ we recall that 
random variables $A_{ij}, \ 1\leq i \leq j \leq n, $ are independent.  Therefore, we can first fix all matrix entries
(up from the diagonal) except $A_{ij} $ and integrate with respect to $A_{ij}.\ $ Applying the Gaussian decoupling formula
\begin{equation}
\label{Gaussdecoupling}
\mbE \eta \* f(\eta) = \sigma^2 \* \mbE f'(\eta), \ \ \eta \sim N(0, \sigma^2),
\end{equation}
with $\eta=A_{ij} $ and $f(\eta)=G_{ij},\ $ and taking into account that $ Var(A_{ij})=\frac{1+\delta_{ij}}{n}, $  and
\begin{equation}
\label{derivatives}
\frac{\partial G_{kl}}{\partial A_{ij}}= \left \{\begin{array} {r@{\quad:\quad}l} 
-G_{ki}\*G_{jl}-G_{kj}\*G_{il} & i \not= j \\
-G_{ki}\*G_{jl} & i=j, \end{array} \right.
\end{equation}
we arrive at
\begin{equation}
\label{mytishchi}
n^{1/3}\* m_{e_1}= n^{2/3} -(2+z\*n^{-2/3})^{-1} \* n^{2/3} -(2+z\*n^{-2/3})^{-1}\*n^{-4/3} \* \mbE \sum_{ij} 
\left( G_{ji}\*G_{ji} + G_{ii}\*G_{jj}\right).
\end{equation}
The term $ n^{-4/3} \* \mbE \sum_{ij} G_{ji}\*G_{ji} $ is equal to
\begin{equation}
\label{schelkovo}
n^{-4/3} \* \mbE \sum_{ij} G_{ji}\*G_{ji}= \mbE n^{-4/3} \* Tr (G_n^2(z))=m_{e_2}. 
\end{equation}
To deal with the term $ n^{-4/3} \* \mbE \sum_{ij} 
G_{ii}\*G_{jj}, \ $ we rewrite it as
\begin{eqnarray}
\label{taininka}
& & n^{-4/3} \* \mbE \sum_{ij} G_{ii}\*G_{jj}= n^{-4/3} \* \mbE \left( (Tr G_n(z))^2\right)= 
n^{-4/3} \* \mbE \left( (-n +n+ Tr G_n(z))^2\right) \\
&=& n^{2/3} - 2\*n^{-1/3} \* \mbE (n+ Tr G_n(z)) + n^{-4/3} \*
\mbE \left( (n+ Tr G_n(z))^2\right) 
=  n^{2/3} -2\*n^{1/3} \* m_{e_1} + m_{2\*e_1}. \nonumber
\end{eqnarray}
As a result, we obtain
\begin{eqnarray}
\label{perlovka}
& & n^{1/3}\* m_{e_1}= n^{2/3} -(2+z\*n^{-2/3})^{-1} \* n^{2/3} -(2+z\*n^{-2/3})^{-1}\*m_{e_2}-(2+z\*n^{-2/3})^{-1}\* n^{2/3}
\nonumber \\
& & +2 \* (2+z\*n^{-2/3})^{-1} \*n^{1/3} \* m_{e_1}-(2+z\*n^{-2/3})^{-1}\*m_{2\*e_1},
\end{eqnarray}
which is equivalent to
\begin{eqnarray}
\label{voronok}
& & n^{2/3} \* (1-(1+z\*n^{-2/3}/2)^{-1} ) - m_{2\*e_1} \*(2+z\*n^{-2/3})^{-1} - m_{e_2} \* (2+z\*n^{-2/3})^{-1}
\nonumber \\
& & =m_{e_1} \*n^{1/3} \*(1-(1+z\*n^{-2/3}/2)^{-1}).
\end{eqnarray}
After trivial arithmetical simplifications, this leads to (\ref{recnule}).  The formula (\ref{recursiveGOE}) can be proven along the same lines if 
one starts with 
$n^{1/3}\*m_{K+e_1}. $ One can say that the formula (\ref{recnule}) gives us the boundary term in the recursive system of 
linear equations satisfied by $ \{m_K \} $ since it corresponds to $K=0. $   Turning our attention to (\ref{recursiveGOE}), we write
\begin{eqnarray}
\label{bolshevo1}
& & n^{1/3}\*m_{K+e_1}= n^{1/3} \* \mbE \left( (g_{n,1}^c(z))^{k_1+1} \* \prod_{l=2}^j  (g_{n,k}(z))^{k_l} \right)=
\nonumber \\
& & \mbE \left[ n^{-1/3} (n + Tr G_n) \* \left( n^{-2/3}\*(n + Tr G_n) \right)^{k_1} \* \prod_{l\geq 2} \left( n^{-2\*l/3} \*
Tr G^l \right)^{k_l} \right],
\end{eqnarray}
we then rewrite, as before, the first term $ n^{-1/3} (n + Tr G_n) $ as
\begin{equation}
\label{pushkino}
n^{-1/3} (n + Tr G_n)=
n^{2/3} -(2+z\*n^{-2/3})^{-1} \* n^{2/3} +(2+z\*n^{-2/3})^{-1}\*n^{-1/3} \* \sum_{ij} A_{ij}\*G_{ji}.
\end{equation}
This leads to
\begin{eqnarray}
\label{los}
& & n^{1/3}\*m_{K+e_1}= n^{2/3}\* m_K -(2+z\*n^{-2/3})^{-1} \* n^{2/3} \*m_K + \nonumber \\
& & (2+z\*n^{-2/3})^{-1}\*n^{-1/3} \* \mbE \left[ (\sum_{ij} A_{ij}\*G_{ji}) \* 
\left( n^{-2/3}\*(n + Tr G_n) \right)^{k_1} \* \prod_{l\geq 2} \left( n^{-2\*l/3} \*
Tr G^l \right)^{k_l} \right].
\end{eqnarray}
As in the case $K=0$ considered above, we fix all matrix entries (up from the diagonal) except $A_{ij},$  and apply 
(\ref{Gaussdecoupling}) with $\eta=A_{ij}$ and 
$f(\eta)= G_{ji} \*\left( n^{-2/3}\*(n + Tr G_n) \right)^{k_1} \* \prod_{l\geq 2} \left( n^{-2\*l/3} \*
Tr G^l \right)^{k_l}. $
Taking into account (\ref{derivatives}) and the equation
\begin{equation}
\label{proizvod}
\frac{\partial Tr (G^l)}{\partial A_{ij}}= -2\*l\* (G^{l+1})_{ij},
\end{equation}
one then obtains (\ref{recursiveGOE}) after some simple algebraic calculations.  Theorem 1.1 is proven.

The proof of Theorem 1.2 is quite similar.  The only alteration required in the GUE case is that 
one needs to replace (\ref{derivatives}) with
\begin{equation}
\label{derivatives1}
\frac{\partial G_{kl}}{\partial Re(A_{ij})}= \left \{\begin{array} {r@{\quad:\quad}l} 
-G_{ki}\*G_{jl}-G_{kj}\*G_{il} & i \not= j \\
-G_{ki}\*G_{jl} & i=j, \end{array} \right.
\end{equation}
and
\begin{equation}
\label{derivatives2}
\frac{\partial G_{kl}}{\partial Im(A_{ij})}= 
- i (G_{ki}\*G_{jl}-G_{kj}\*G_{il}) \ \ \text{for} \ i \not= j.
\end{equation}
The remaining calculations are very similar and are left to the reader.

\section{Proof of Theorems 1.3 and 1.4}

The proofs will be similar to the ones given in the previous section. 
Let us start with the proof of Theorem 1.3.  Our first goal is to establish (\ref{recnule3}).
To this end, we consider $n^{-1}\* m_{e_1}= \mbE \left( n^{-3}\* Tr G_{n, N}(t)  \right), $ where
$G_{n,N}(t) $ was defined in (\ref{lgov}).
By using the resolvent identity
\begin{equation}
\label{rezolventy11}
G_{n,N}=(M_{n,N} +\frac{t^2}{n^2})^{-1}=
\frac{n^2}{t^2} \*Id - \frac{n^2}{t^2}\* A\*A^t\* G_{n,N}
\end{equation}
we arrive at
\begin{equation}
\label{moskva11}
n^{-1}\* m_{e_1}= \frac{1}{t^2} - \frac{1}{n}\*\frac{1}{t^2} \* \mbE \sum_{1\leq i,j \leq n}
\sum_{1\leq p \leq N} A_{ip}\* A_{jp} \*G_{ji}.
\end{equation}
Here $A_{ip}=\frac{a_{ip}}{\sqrt{n}}$ denote the matrix entries of $A_n$, and $G_{ji}$ denote the matrix entries of $G_{n,N}(z).$
To calculate  $\mbE  A_{ip}\*A_{jp}\*G_{ji}, $ we again use 
the Gaussian decoupling formula (\ref{Gaussdecoupling}) and the equation
\begin{equation}
\label{derivatives3}
\frac{\partial G_{kl}}{\partial A_{ip}}=  
-G_{ki}\*(A^t\*G)_{pl}-(G\*A)_{kp}\*G_{il} .
\end{equation}
Therefore,
\begin{eqnarray}
\label{piter11}
& & n^{-1}\* m_{e_1}= \frac{1}{t^2} + \frac{1}{n^2}\*\frac{1}{t^2}\* \mbE \sum_{1\leq i,j \leq n}  \* \sum_{1\leq p \leq N}
A_{jp} \* \left( G_{ji}\* (G\*A)_{ip} + G_{ii}\* (G\*A)_{jp} \right) - \frac{1}{n^2}\*\frac{1}{t^2} 
\* \mbE \* \sum_{1\leq i\leq n} \*  \sum_{1\leq p \leq N}G_{ii}
\cr
& &= \frac{1}{t^2}+  \frac{1}{n^2}\*\frac{1}{t^2}\* \mbE \left[ Tr (G\*A\*A^t\*G) + Tr(G\*A\*A^t)\*Tr G \right] 
-\frac{N}{n^2}\* \frac{1}{t^2} \* \mbE Tr G.
\end{eqnarray}
Using the identity $ G\*A\*A^t= Id -\frac{t^2}{n^2}\* G, \ $ the last formula can  rewritten as
\begin{equation}
\label{davydkovo}
 n^{-1}\* m_{e_1}=\frac{1}{t^2}+ \frac{1}{t^2}\* m_{e_1} - m_{e_2} + \frac{n}{t^2}\* m_{e_1} - m_{2\*e_1} - \frac{n+\nu}{t^2}
\* m_{e_1},
\end{equation}
which implies (\ref{recnule3}).  The formula (\ref{recursiveWishartreal}) can be proven along the same lines if one starts
with $n^{-1} \* m_{K+e_1}. $   Let us write
\begin{eqnarray}
\label{bolshevo}
& & n^{-1}\*m_{K+e_1}= n^{-1} \* \mbE \left( (g_{1}(t))\* \prod_{l=1}^j  (g_{l}(t))^{k_l} \right)=
\nonumber \\
& & \mbE \left[ n^{-3}\*  Tr G_{n,N}) \* \prod_{l\geq 1} \left( n^{-2\*l} \*
Tr G^l_{n,N} \right)^{k_l} \right],
\end{eqnarray}
Using the resolvent identity, we can rewrite the first term in the product as 
$$ G_{n,N}= \frac{n^2}{t^2} \*Id - \frac{n^2}{t^2}\* A\*A^t\* G_{n,N}. $$ After integration by parts and a few lines of 
careful calculations, we obtain (\ref{recursiveWishartreal}).   Theorem 1.4 can be proven along similar lines.

\section{Non-Gaussian Case.}

The generalization of the Gaussian decoupling formula (\ref{Gaussdecoupling}) to the non-Gaussian case 
can be found, for example, in \cite{KKP}:
\begin{equation}
\label{nonGaussdec}
\mbE [\xi\*f(\xi) ]= \sum_{k=0}^p \frac{c_{k+1}}{k!}\* \mbE\left[\frac{d^kf}{dx^k}(\xi)\right] + \varepsilon,
\end{equation}
where $\xi $ is a real random variable such that $ \mbE \left(|\xi|^{p+2} \right) <\infty, \ \ c_l, \ l\geq 1, $
are the cumulants of the random variable $\xi, \ $ complex-valued function
$f(x) $ has first $p+1$ continuous and bounded derivatives, and the error term satisfies the upper bound
$|\varepsilon| \leq B_k\* sup_x |\frac{d^kf}{dx^k}(x)| \* \mbE \left(|\xi|^{p+2} \right) \ $ with the constant 
$B_k $ depending only on $k.$  

It is conjectured that the distribution of the largest eigenvalues in Wigner random matrices is universal provided the fourth moment of the matrix 
entries is finite.  Currently, we are unable to prove this conjecture.  Instead, we speculate below on the possible approach to extend the results 
of Theorems 1.1-1.4 to the non-Gaussian case.   Let us consider a real Wigner random matrix $A_n =\frac{1}{\sqrt{n}}\* (a_{ij})_{i,j=1}^n, \ $
and assume that the entries $(a_{ij}=a_{ji})_{i<j} \ $ are i.i.d. centralized random variables with the unit variance and the finite fourth moment.  
In addition, we assume that the diagonal entries $ a_{ii}, \ 1\leq i \leq n, $ are i.i.d. centralized random variables, independent from the 
non-diagonal entries.  We assume that the diagonal entries also have the finite fourth moment.  Let us also assume for simplicity that
$Var (a_{ii}) =2. \ $  In an attempt to extend the result of Theorem 1.1 to the non-Gaussian situation, we apply the generalized decoupling formula
(\ref{nonGaussdec}).  To be specific, let us concentrate our attention  on the ``boundary'' equation (\ref{recnule}).   Looking at  (\ref{moskva}),
we apply (\ref{nonGaussdec}) to $ \mbE A_{ij}\*G_{ji}. \ $ Since $c_1(a_{ij})=0, \ c_2(a_{ij})=1+\delta_{ij}, \ 
c_3(a_{ij})=c_3, \ $ for $i<j, \ $ and $ \mbE a_{ij}^4 \leq \infty, \ $ one might wish to truncate (\ref{nonGaussdec}) after the first three terms (i.e.
$p=2$) to obtain
\begin{equation}
\label{museumplein}
n ^{-1/3} \* \mbE \sum_{ij} A_{ij}\*G_{ji}= -n^{-4/3} \* \mbE \sum_{ij} (G^2_{ji} + G_{ii}\*G_{jj}) + 
n^{-11/6} \* \mbE \sum_{ij} (3\*G_{ij}\*G_{ii}\*G_{jj} + G^3_{ij})  +\epsilon.
\end{equation}
The first sum in (\ref{museumplein}) is the same as in the Gaussian case.  The hope is to show that the second sum and the remainder term give 
negligible contributions in the limit $n \to \infty. $   However, it is currently unclear to us how to efficiently bound the terms
$ n^{-11/6} \* \mbE \sum_{ij} G_{ij}\*G_{ii}\*G_{jj} $ and $ n^{-11/6} \* \mbE \sum_{ij} G^3_{ij}. $

\end{document}